\documentclass[aps,prd,notitlepage,nofootinbib,showpacs]{revtex4-1}

\usepackage{graphicx}
\usepackage{amssymb}
\usepackage{amsmath}
\usepackage{bm}
\usepackage{latexsym}
\usepackage{epsfig}
\usepackage{textcomp}
\usepackage{color}

\def\nn{\nonumber} 

\def\f{\frac}
\def\l{\left}
\def\r{\right}
\def\d{{\rm d}}
\def\Mpl{M_{_{\rm Pl}}}
\def\beq{\begin{equation}}
\def\eeq{\end{equation}} 
\def\beqa{\begin{eqnarray}}
\def\eeqa{\end{eqnarray}} 

\def\bA{\bar A}
\def\cA{\mathcal A}

\def\psb{{\mathcal P}_{_{\rm B}}}
\def\pse{{\mathcal P}_{_{\rm E}}}

\newcommand{\viz}{\textit{viz.~}}
\newcommand{\ie}{\textit{i.e.~}}

\begin{document}

\title{Duality and scale invariant magnetic fields from bouncing universes}
\author{Debika Chowdhury}
\email{debika@physics.iitm.ac.in}
\affiliation{Department of Physics, Indian Institute of Technology Madras, 
Chennai~600036, India}
\author{L.~Sriramkumar}
\email{sriram@physics.iitm.ac.in}
\affiliation{Department of Physics, Indian Institute of Technology Madras, 
Chennai~600036, India}
\author{Rajeev Kumar Jain} 
\affiliation{CP$^3$-Origins, Centre for Cosmology and Particle Physics 
Phenomenology 
University of Southern Denmark, Campusvej 55, 5230 Odense M, Denmark}
\email{jain@cp3.sdu.dk}


\begin{abstract}
Recently, we had numerically shown that, for a non-minimal coupling
that is a simple power of the scale factor, scale invariant magnetic fields
arise in a class of bouncing universes.
In this work, we {\it analytically}\/ evaluate the spectrum of magnetic and 
electric fields generated in a sub-class of such models.
We illustrate that, for cosmological scales which have wavenumbers much smaller
than the wavenumber associated with the bounce, the shape of the spectrum is 
preserved across the bounce.
Using the analytic solutions obtained, we also illustrate that the problem of
backreaction is severe at the bounce.
Finally, we show that the power spectrum of the magnetic field remains invariant 
under a two parameter family of transformations of the non-minimal coupling 
function.
\end{abstract}


\maketitle


\section{Introduction}

Magnetic fields are ubiquitous in the universe.
Coherent magnetic fields have been observed over a wide variety of scales,
ranging from astrophysical systems such as stars and galaxies, to
cosmological systems such as the large scale structures (in this context,
see Refs.~\cite{Grasso:2000wj,Widrow:2002ud}; for some recent reviews, see
Refs.~\cite{Kandus:2010nw,Widrow:2011hs,Durrer:2013pga,Subramanian:2015lua}).
More recently, magnetic fields have been observed even in the intergalactic medium~\cite{Neronov:1900zz,Tavecchio:2010mk}.
While the strength of magnetic fields observed in galaxies and clusters of
galaxies are typically about a few micro Gauss, in the intergalactic medium,
the lower bounds on their strengths have been inferred to be of the order of $10^{-17}$ Gauss at $1$ Mpc from the Fermi/LAT and HESS observations of TeV blazars
(see Refs.~\cite{Neronov:1900zz,Tavecchio:2010mk,Dermer:2010mm,Vovk:2011aa};
also see Refs.~\cite{Tavecchio:2010ja,Dolag:2010ni,Taylor:2011bn,Takahashi:2011ac,
Huan:2011kp,Finke:2015ona}).
This should be contrasted with the upper bound of a few nano Gauss which has
been arrived at from the CMB observations (for current constraints from the
PLANCK and POLARBEAR data, see Refs.~\cite{Ade:2015cva,Ade:2015cao} and
references therein).
Similar upper limits have also been obtained independently using the rotation
measures from the NRAO VLA Sky Survey~\cite{Pshirkov:2015tua}.
These bounds are broadly in agreement with the limits arrived at from the LSS
data, either alone or when combined with the CMB data (in this context, see
Refs.~\cite{Yamazaki:2010nf,Kahniashvili:2010wm,Caprini:2011cw}; for improved
limits from LSS and reionization, see Refs.~\cite{Shaw:2010ea,Schleicher:2011jj,
Fedeli:2012rr,Kahniashvili:2012dy,Pandey:2014vga}.
Though astrophysical processes such as the dynamo mechanism can, in principle,
boost the amplitude of magnetic fields in galaxies, a seed field is nevertheless
required for such mechanisms to work.
Therefore, a primordial origin for the magnetic fields seems inevitable to
explain their prevalence, particularly on the largest scales.

\par

Inflation is currently considered the most promising paradigm to describe 
the origin of perturbations in the early universe.
Hence, it seems natural to consider the generation of magnetic fields in 
the inflationary scenario.
It is well known that the conformal invariance of the electromagnetic field 
has to be broken in order to generate magnetic fields of observable strengths
in the early universe~\cite{Turner:1987bw, Ratra:1991bn}. 
In fact, the issue has been studied extensively.
There exist many inflationary models which lead to nearly scale invariant 
magnetic fields of appropriate strength and correlation scales to match with 
the observations~\cite{Bamba:2003av,Bamba:2006ga,Demozzi:2009fu,Martin:2007ue,
Campanelli:2008kh,Kanno:2009ei,Subramanian:2009fu,Urban:2011bu,Durrer:2010mq,
Byrnes:2011aa,Jain:2012jy, Kahniashvili:2012vt,Cheng:2014kga,Bamba:2014vda,Fujita:2015iga,Campanelli:2015jfa,
Fujita:2016qab,Tsagas:2016fax}.   
However, most models of inflationary magnetogenesis typically suffer from the 
so-called backreaction and strong coupling problems (see, for instance, 
Refs.~\cite{Demozzi:2009fu,Ferreira:2013sqa,Ferreira:2014hma}).

\par

Under such circumstances, it seems worthwhile to examine the generation of 
magnetic fields in alternative scenarios of the early universe.
A reasonably popular alternative are bouncing models wherein the universe 
undergoes a period of contraction until the scale factor attains a minimum 
value, after which it begins to expand (see, for instance, 
Refs.~\cite{Finelli:2001sr,Martin:2001ue,Tsujikawa:2002qc,Peter:2002cn,
Martin:2003sf,Martin:2003bp, Peter:2003rg, Martin:2004pm,Allen:2004vz,
Battefeld:2004mn,Creminelli:2004jg,Geshnizjani:2005hc,Abramo:2007mp,
Peter:2008qz,Falciano:2008gt,Cardoso:2008gz}, and the following 
reviews~\cite{Novello:2008ra,Battefeld:2014uga,Brandenberger:2016vhg,Cai:2014bea}). 
Such bouncing scenarios provide an alternative to inflation to overcome the 
horizon problem. 
These models allow well motivated, Minkowski-like initial conditions to be 
imposed on the perturbations at early times during the contracting 
phase.
The generation of magnetic fields in such scenarios have been explored only
to a limited extent~\cite{Salim:2006nw,Membiela:2013cea,Sriramkumar:2015yza}.

\par

In a recent work~\cite{Sriramkumar:2015yza}, we had numerically shown that,
for non-minimal couplings that are a simple power of the scale factor, scale 
invariant magnetic fields can be generated in certain bouncing scenarios.
In this work, we investigate the problem analytically in a sub-class of these
models wherein the non-minimal coupling is a positive power of the scale factor.
We consider a specific form for the scale factor, leading to a non-singular 
bounce, which reduces to a power law form far from the bounce.
We find that, in such situations, we can obtain analytical solutions for modes 
of the electromagnetic vector potential that are much smaller than the natural 
scale associated with the bounce.   
We divide the time before the bounce into two domains, one that corresponds 
to very early times and another closer to the bounce. 
We analytically evaluate the electromagnetic modes during these domains and 
arrive at the corresponding power spectra for the electric and magnetic fields. 
It can be easily shown that scale invariant magnetic fields can be generated 
before the bounce for specific values of the parameters involved. 
We evolve these modes across the bounce and calculate the power spectra in the 
final and third domain, \ie\/ in the early stages of the expanding phase 
{\it after}\/ the bounce.
We show that the shapes of the spectra are preserved for scales of cosmological
interest as the modes evolve across the bounce.

\par

At this stage, it is important that we comment on the 
theoretical and observational status of bouncing models.
Theoretically, the main issue that plagues these models is the rapid growth 
of perturbations as one approaches the bounce.
Evidently, this raises questions about the validity of linear perturbation 
theory around the bounce~\cite{Battefeld:2014uga,Brandenberger:2016vhg}.
As far as scalar perturbations are concerned, this is typically circumvented
by working in a specific gauge where the amplitude of the perturbations remain
small (in this context, see, for instance, Ref.~\cite{Allen:2004vz}).
Another concern that had been pointed out early in the literature is the rapid 
growth of vector perturbations in a contracting universe~\cite{Battefeld:2004cd}.
But, this issue does not arise if one assumes that there are no vector sources.
As far as the observational constraints are concerned, one finds that nearly 
scale invariant scalar and tensor power spectra can indeed be generated in 
bouncing scenarios~\cite{Allen:2004vz,Finelli:2007tr}.
However, one of the primary problems that confronts bouncing models seems to be 
the fact that the tensor-to-scalar ratio $r$ generated in these models may prove
to be much larger than the present upper bound of $r \lesssim 0.1$ from 
Planck~\cite{Ade:2015lrj}.
For instance, in certain matter bounce scenarios, the tensor-to-scalar ratio $r$
has been found to be as large as ${\cal O}(10)$~\cite{Allen:2004vz,Cai:2014xxa,
Battefeld:2014uga,Cai:2008qw,Cai:2014bea}, which is considerably beyond the
constraints arrived at from the CMB observations.
Nonetheless, these exist other models---such as the matter bounce curvaton 
scenario~\cite{Cai:2011zx} and other versions of the matter bounce 
scenario~\cite{Cai:2012va,Cai:2013kja}---which lead to values of $r$ that seem
to be consistent with the observations.

\par

This paper is organized as follows. 
In the following section, we shall describe a few essential aspects of the 
electromagnetic field that is coupled non-minimally to a scalar field.
We shall discuss the equation of motion governing the electromagnetic potential,
the quantization of the potential in terms of the normal modes in an evolving
universe and the power spectra describing the electric and magnetic fields. 
We shall also introduce the forms of the scale factor and the coupling function
that we consider. 
In Sec.~\ref{sec:ps}, we shall divide the bounce into three domains and evaluate 
the modes analytically in each of these domains. 
We shall evaluate the power spectra prior to the bounce as well as soon after 
the bounce and illustrate that the shape of the power spectra are preserved
across the bounce. 
In Sec.~\ref{sec:br}, we shall study the issue of backreaction using the analytic 
solutions for the modes. 
In Sec.~\ref{sec:sidi}, we shall illustrate that the power spectrum of 
the magnetic field is form invariant under a two parameter family of 
transformations of the coupling function.
Finally, we shall conclude with a brief discussion in Sec.~\ref{sec:dis}.

\par

We shall work with natural units such that $\hbar=c=1$, and set the Planck mass 
to be $\Mpl=\l(8\,\pi\, G\r)^{-1/2}$. 
We shall adopt the metric signature of $\l(-, +, +, +\r)$. 
Greek indices shall denote the spacetime coordinates, whereas the Latin indices 
shall represent the spatial coordinates, except for $k$ which shall be reserved 
for denoting the wavenumber. 
Lastly, an overprime shall denote differentiation with respect to the conformal 
time coordinate.


\section{The bounce, non-minimal action, equations of motion and 
power spectra}\label{sec:nma}

Recall that, if $A^{\mu}$ is the electromagnetic vector potential, then
the corresponding field tensor $F_{\mu\nu}$ is given in terms of $A^{\mu}$ 
by the relation
\begin{equation}
F_{\mu\nu} = A_{\nu;\mu}-A_{\mu;\nu}= A_{\nu,\mu}-A_{\mu,\nu}.
\end{equation}
We shall consider the case wherein the electromagnetic field is coupled 
non-minimally to a scalar field $\phi$ through a function $J(\phi)$ and 
is described by the action (see, for instance, Refs.~\cite{Bamba:2006ga,
Martin:2007ue,Subramanian:2009fu})
\begin{equation}\label{eq:nm-em}
S[\phi,A^{\mu}]
=-\frac{1}{16\,\pi} \int \d^{4}x\, \sqrt{-g}\, J^2(\phi)\, F_{\mu\nu}F^{\mu\nu}.
\end{equation}
Evidently, it is the coupling function $J$ which is responsible for 
breaking the conformal invariance of the action.
The scalar field $\phi$, for example, can be the primary source that is 
driving the evolution of the bouncing model.
The variation of the above action leads to the following equation of motion of 
the electromagnetic field
\begin{equation}
\frac{1}{\sqrt{-g}}\;{\partial}_{\mu} \l[\sqrt{-g}\, J^2(\phi)\, F^{\mu\nu}\r]=0.
\label{eq:eom}
\end{equation}

\par

We shall consider the background to be the spatially flat, 
Friedmann-Lema\^itre-Robertson-Walker (FLRW) metric that is described by 
the line-element
\begin{equation}
\d s^2 = a^2(\eta)\, 
\l(-\d\eta^2+\delta_{ij}\, \d x^i\,\d x^j\r),
\end{equation}
where $a(\eta)$ is the scale factor and $\eta$ denotes the conformal time 
coordinate.
In order to study the evolution of the vector potential, we shall choose to 
work in the Coulomb gauge wherein $A_0 = 0$ and $\partial_i\,A^i = 0$.
On quantization, the vector potential $\hat{A_i}$ can be Fourier decomposed as
follows~\cite{Martin:2007ue,Subramanian:2009fu,Ferreira:2013sqa}:
\begin{equation}
\hat{A_i}\l(\eta,{\bm x}\r) 
=\sqrt{4\,\pi}\int\frac{\d^3\,{\bm k}}{(2\,\pi)^{3/2}}\,
\sum_{\lambda=1}^2\tilde{\epsilon}_{\lambda i}({\bm k})\,
\l[\hat{b}_{\bm k}^{\lambda}\, {\bar A}_k(\eta)\,
{\rm e}^{i\,{\bm k}\cdot{\bm x}}
+ \hat{b}_{\bm k}^{\lambda\dagger}\, {\bar A}_k^{\ast}(\eta)\,
{\rm e}^{-i\,{\bm k}\cdot{\bm x}}\r],\label{eq:vp-d}
\end{equation}
where the Fourier modes ${\bar A}_{k}$ satisfy the differential equation
[cf.~Eq.~(\ref{eq:eom})]
\begin{equation}
{\bar A}_k''+2\,\f{J'}{J}\,{\bar A}_k'+k^2\, {\bar A}_k=0.
\label{eq:de-Abk-o}
\end{equation}
The quantities $\tilde{\epsilon}_{\lambda i}$ represent polarization 
vectors and the summation corresponds to the two orthonormal transverse 
polarizations. 
The operators $\hat{b}_{\bm k}^{\lambda}$ and 
$\hat{b}_{\bm k}^{\lambda\dagger}$ are the annihilation and creation 
operators satisfying the following standard commutation relations:
\begin{equation}
\l[\hat{b}_{\bm k}^{\lambda},
\hat{b}_{\bm k'}^{\lambda'}\r] 
= \l[\hat{b}_{\bm k}^{\lambda\dagger},
\hat{b}_{\bm k'}^{\lambda'\dagger}\r] = 0,\quad
\l[\hat{b}_{\bm k}^{\lambda},
\hat{b}_{\bm k'}^{\lambda'\dagger}\r] 
=\delta_{\lambda\lambda'}\;
\delta^{(3)}\l({\bm k} - {\bm k'}\r).
\end{equation}
Let us now define a new variable ${\mathcal A}_k=J\, {\bar A}_k$, which, as 
we shall see, proves to be convenient to deal with.
In terms of the new variable, Eq.~(\ref{eq:de-Abk-o}) for ${\bar A}_k$ 
simplifies to 
\begin{equation}
{\mathcal A}_k''+\l(k^2-\f{J''}{J}\r)\,{\mathcal A}_k=0.\label{eq:de-cAk}
\end{equation}

\par

Let $\hat{\rho}_{_{\rm E}}$ and $\hat{\rho}_{_{\rm B}}$ denote the operators 
corresponding to the energy densities associated with the electric and magnetic 
fields.
Upon using the decomposition~(\ref{eq:vp-d}) of the vector potential, the 
expectation values of the energy densities $\hat{\rho}_{_{\rm E}}$ and 
$\hat{\rho}_{_{\rm B}}$ can be evaluated in the vacuum state, say, $\vert 
0\rangle$, that is annihilated by the operator ${\hat b}_{\bm k}^{\lambda}$.
It can be shown that the spectral energy densities of the magnetic and electric 
fields can be expressed in terms of the modes ${\bar A}_k$ and $\cA_k$, their 
derivatives ${\bar A}_k'$ and $\cA_k'$, and the coupling function $J$ as 
follows~\cite{Martin:2007ue,Subramanian:2009fu}:
\begin{subequations}
\label{eq:ps}
\begin{eqnarray}
\psb(k)
&=&\frac{\d\langle 0 \vert \hat{\rho}_{_{\rm B}}\vert 0 \rangle}
{\d\,{\rm ln}\,k}\nn\\
&=&\f{J^2(\eta)}{2\,\pi^{2}}\,\f{k^{5}}{a^{4}(\eta)}\,
\vert{\bar A}_k(\eta)\vert^{2}
=\f{1}{2\,\pi^{2}}\,\f{k^{5}}{a^{4}(\eta)}\,
\vert \cA_k(\eta)\vert^{2},\label{eq:psb}\\
\pse(k)&=&\frac{\d\langle 0\vert \hat{\rho}_{_{\rm E}}\vert 0 \rangle}
{\d\,{\rm ln}\,k}\nn\\
&=&\f{J^{2}(\eta)}{2\,\pi^{2}}\,\frac{k^{3}}{a^{4}(\eta)}\,
\vert{\bar A}_k'(\eta)\vert^2
=\f{1}{2\,\pi^{2}}\,\frac{k^{3}}{a^{4}(\eta)}\;
\biggl\vert\cA_k'(\eta)
-\f{J'(\eta)}{J(\eta)}\,\cA_k(\eta)\biggr\vert^2.\label{eq:pse}
\end{eqnarray}
\end{subequations}
The spectral energy densities $\psb(k)$ and $\pse(k)$ are often referred to as 
the power spectra for the generated magnetic and electric fields, respectively. 
A flat or scale invariant magnetic field spectrum corresponds to a constant, 
\ie\/ $k$-independent, ${\mathcal P}_{_{\rm B}}(k)$.

\par

We shall model the non-singular bounce by assuming that the scale factor 
$a(\eta)$ behaves as follows~\cite{Sriramkumar:2015yza,Chowdhury:2015cma}:
\begin{equation}
a(\eta)=a_{0}\, \l(1+\f{\eta^2}{\eta_{0}^{2}}\r)^{q}
= a_{0}\, \l(1+k_0^2\,\eta^2\r)^{q},\label{eq:sf}
\end{equation}
where $a_{0}$ is the value of the scale factor at the bounce (\ie\/ when $\eta=0$), 
$\eta_{0}=1/k_0$ denotes the time scale of the duration of the bounce and $q>0$. 
Note that, when $q=1$, during very early times wherein $\eta \ll -\eta_0$, the 
scale factor behaves as in a matter dominated universe (\ie\/ $a \propto \eta^2$). 
Therefore, the $q=1$ case is often referred to as the matter bounce scenario. 
We shall assume that the scale $k_0$ associated with the bounce is of the order 
of the Planck scale~$\Mpl$.
We should mention here that, for certain values of the parameters 
involved, the above scale factor leads to tensor power spectra that are
consistent with the CMB observations (see, for instance, 
Refs.~\cite{Chowdhury:2015cma,Stargen:2016cft}).
However, we should hasten to add that determining the corresponding scalar power 
spectra requires a detailed modeling of the bounce.

\par

The scale factor~(\ref{eq:sf}) above can be achieved, for 
instance, if we consider that the universe is composed of two non-interacting 
fluids with constant equation of state parameters. 
Let the energy densities and pressure of the two fluids be denoted by $\rho_i$ 
and $p_i$ respectively, with $i = (1,2)$. 
Also, let the equations of state for the two fluids be given by $p_i = w_i\,
\rho_i$, where $w_i$ is a constant.
Since the two fluids are non-interacting, the equation governing the conservation 
of energy associated with the fluids can be integrated to yield that $\rho_i = M_i
/a^{r_i}$, where $M_i$ is a constant.
As is well known, the index $r_i$ is related to the equation of state parameter 
$w_i$ by the relation: $r_i = 3\,\l(1 + w_i\r)$.
It can be easily shown that the equation of state parameters $w_1$ and $w_2$
are related to the quantity $q$ through the relations:~$w_1=(1-q)/(3\,q)$ and
$w_2=(2-q)/(3\,q)$.  
Further, one can show that $M_1=12\,k_0^2\,\Mpl^2\, a_0^{1/q}$ and $M_2=-M_1\,
a_0^{1/q}$.
It is important to note that, while $M_1$ is positive, $M_2$ is negative.
In other words, the energy density of the second fluid is always negative. 
This seems inevitable as the total energy density has to vanish at the bounce. 
For the specific case of $q=1$, which is the matter bounce scenario, the first 
fluid corresponds to matter.
The second fluid behaves in a manner similar to radiation as far as its time 
evolution is concerned, but it has a negative energy density.
For our discussion, we shall assume that the evolution of the universe is
achieved with the aid of suitable scalar field(s) which effectively mimic 
the behavior of the fluids (in this context, see, for example, 
Ref.~\cite{Unnikrishnan:2010ncf}).

\par

Given a scale factor, in order to arrive at the behavior of the electromagnetic
modes in a FLRW universe, we shall also require the form of the non-minimal
coupling function $J$.
We shall assume that the coupling function can be conveniently expressed in terms 
of the scale factor as follows:
\begin{equation}
J(\eta)=J_0\, a^n(\eta).\label{eq:J}
\end{equation}
It can be easily argued that the resulting power spectra are independent of 
the constant $J_0$ (in this context, see Ref.~\cite{Sriramkumar:2015yza}).
As we shall discuss in the following section, in this work, for the problem 
to be tractable completely analytically, we shall restrict ourselves to cases 
wherein $n$ is positive.  


\section{Analytical evaluation of the modes and the power spectra}\label{sec:ps}

To arrive at analytic solutions to the electromagnetic modes, let us divide the 
period prior to the bounce into two domains, one far away from the bounce and 
another closer to the bounce.
Let these two domains correspond to $-\infty<\eta<-\alpha\,\eta_0$ and $-\alpha\,
\eta_0 <\eta<0$, where $\alpha$ is a relatively large number, say, of the order 
of $10^5$ or so.

\par

During the first domain, the scale factor~(\ref{eq:sf}) reduces to the 
following power law form:~$a(\eta)\propto \eta^{2\,q}$.
In such a case, the non-minimal coupling function $J$ also simplifies to a 
power law form and it behaves as $J(\eta) \propto \eta^\gamma$, where we 
have set $\gamma =2\,n\,q$.
Under these conditions, we have $J''/J\simeq \gamma\,(\gamma-1)/\eta^2$.
This behavior is exactly what is encountered for a similar coupling function
in power law inflation.
Due to this reason, it is straightforward to show that the solutions to the 
modes of the electromagnetic vector potential $\bA_k$ in the first domain can 
be expressed in terms of the Bessel functions $J_\nu(x)$~\cite{Martin:2007ue,
Subramanian:2009fu,Ferreira:2013sqa,Membiela:2013cea,Sriramkumar:2015yza}.
One finds that the solutions can be expressed in terms of the quantity $\cA_k$ 
as follows:
\begin{equation}
\cA_k(\eta)  
= \sqrt{-k\, \eta}\, \l[C_{1}(k)\, J_{\gamma-1/2}(-k\, \eta) 
+ C_{2}(k)\, J_{-\gamma + 1/2} (-k\,\eta)\r],\label{eq:cA-d1}
\end{equation}
where the coefficients $C_{1}(k)$ and $C_{2}(k)$ are to be fixed by the initial 
conditions. 
On imposing the Bunch-Davies initial conditions at early times during the 
contracting phase, \ie\/ as $k\,\eta\to-\infty$, one obtains that 
\begin{eqnarray}
C_1(k)
=\sqrt{\f{\pi}{4\,k}}\; 
\f{{\rm e}^{-i\,\pi\,\gamma/2}}{{\rm cos}(\pi\,\gamma)}, \qquad
C_2(k)
=\sqrt{\f{\pi}{4\,k}}\; 
\f{{\rm e}^{i\,\pi\,(\gamma+1)/2}}{{\rm cos}(\pi\,\gamma)}.
\end{eqnarray}
At this stage, it is also useful to note that 
\begin{equation}
\cA_k'(\eta)-\f{J'}{J}\,\cA_k(\eta)   
= k\,\sqrt{-k\, \eta}\, \l[C_{1}(k)\, J_{\gamma+1/2}(-k\, \eta) 
- C_{2}(k)\, J_{-\gamma - 1/2} (-k\,\eta)\r],\label{eq:cAp-d1}
\end{equation}
an expression we shall require to evaluate the power spectrum of the electric
field.

\par

\begin{figure}[!t]
\begin{center}
\includegraphics[width=12.0cm]{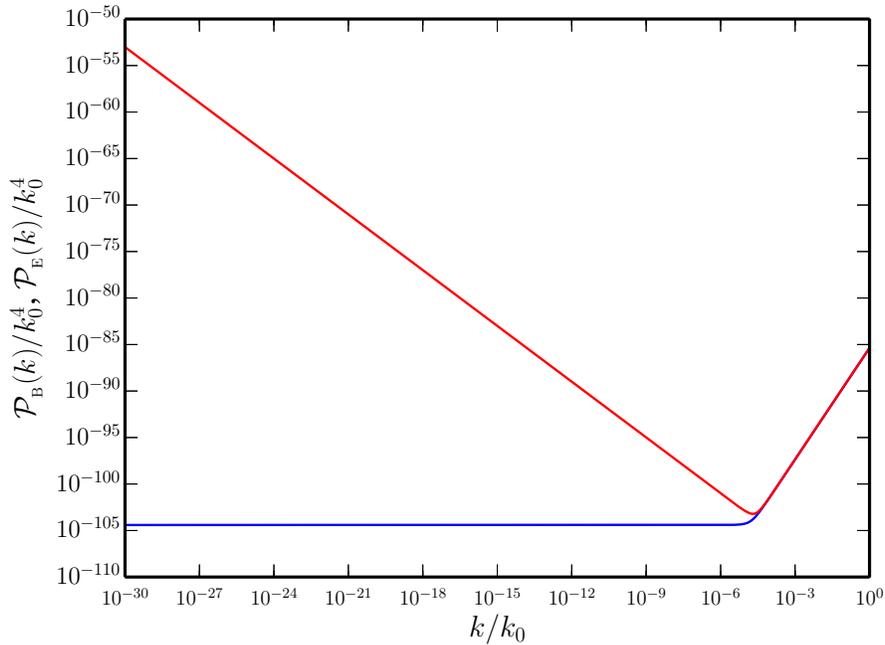}
\end{center}
\caption{The power spectra of the magnetic (in blue) and electric (in red) 
fields, evaluated before the bounce at $\eta=-\alpha\,\eta_0$ using the 
analytical expressions~(\ref{eq:cA-d1}) and~(\ref{eq:cAp-d1}), have been 
plotted as a function of $k/k_0$ for $\gamma=3$, $q=1$, $a_0 = 8.73\times 
10^{10}$ and $\alpha=10^5$.
Note that the dimensionless quantities ${\cal P}_{_{\rm B}}(k) /k_0^4$ and 
${\cal P}_{_{\rm E}}(k)/k_0^4$ that we have plotted depend only on the
combination $k/k_0$.
Also, we should mention that our analytical approximations are valid only
for scales such that $k\ll k_0/\alpha$.
Over this domain, while the spectrum of the magnetic field is strictly scale 
invariant, the spectrum of the electric field behaves as $k^{-2}$, as is 
suggested by the spectra~(\ref{eq:psb-bb}) and~(\ref{eq:pse-bb}) arrived at 
from the asymptotic forms of the Bessel functions.
Needless to add, the question of interest is whether these power spectra will
retain their shape after the bounce.}
\label{fig:ps-bb}
\end{figure}

Let us now evaluate the power spectra of the magnetic and electric fields as 
one approaches the bounce, \ie\/ in the limit $k\,\vert \eta\vert \ll 1$.
It should be mentioned that, in order for the solutions we have obtained above 
to be applicable, we need to remain in the first domain (\ie\/ $-\infty<\eta<
-\alpha\,\eta_0$) even as we consider this limit.
This condition implies that we have to restrict ourselves to modes such that
$k\ll k_0/\alpha$. 
The power spectra of the magnetic and electric fields can be arrived at from 
the above expressions for $\cA_k$ and  $\cA_k'-(J'/J)\,\cA_k$ and the asymptotic 
forms of the Bessel functions.
As is to be expected, the resulting spectra have the same form as one 
encounters in power law inflation.  
One finds that the spectrum of the magnetic field can be written 
as~\cite{Martin:2007ue,Subramanian:2009fu,Ferreira:2013sqa,Sriramkumar:2015yza}
\begin{equation}
{\cal P}_{_{\rm B}}(k) 
= \frac{{\cal F}(m)}{2\,\pi^2}\, \l(\f{H}{2\, q}\r)^4 (-k\,\eta)^{4+2\,m},
\label{eq:psb-bb}
\end{equation}
where $H\simeq (2\,q/a_0\,\eta)\, (\eta_0/\eta)^{2\,q}$, while $m=\gamma$ for 
$\gamma \le 1/2$ and $m =1-\gamma$ for $\gamma \ge 1/2$.
Moreover, the quantity ${\cal F}(m)$ is given by
\begin{equation}
{\cal F}(m) = \frac{\pi}{2^{2\,m+1}\,\Gamma^2(m+1/2)\,\cos^2(\pi\,m)}.
\end{equation}
Clearly, the case $m=-2$ leads to a scale invariant spectrum for the magnetic 
field, which corresponds to either $\gamma=3$ or $\gamma=-2$.
The associated spectrum for the electric field can be evaluated to be
\begin{equation}
{\cal P}_{_{\rm E}}(k) 
= \frac{{\cal G}(m)}{2\,\pi^2}\, \l(\f{H}{2\,q}\r)^4 (-k\,\eta)^{4+2\,m},
\label{eq:pse-bb}
\end{equation}
where $m=1+\gamma$ if $\gamma \le -1/2$ and $m =-\gamma$ for $\gamma \ge -1/2$, 
while ${\cal G}(m)$ is given by
\begin{equation}
{\cal G}(m) = \frac{\pi}{2^{2\,m+3}\, \Gamma^2(m +3/2)\,\cos^2(\pi\,m)}. 
\end{equation}
It should be noted that, when $\gamma=3$ and $\gamma=-2$, the power 
spectrum of the electric field behaves as $k^{-2}$ and $k^2$, respectively.
These results imply that, in the bouncing scenario, one can expect these 
cases to lead to scale invariant spectra (corresponding to wavenumbers 
such that $k\ll k_0/\alpha$) for the magnetic field before the bounce.
Using the analytic expressions~(\ref{eq:cA-d1}) and~(\ref{eq:cAp-d1}), 
in Fig.~\ref{fig:ps-bb}, we have plotted the spectra of the magnetic 
and electric fields evaluated at $\eta=-\alpha\,\eta_0$ as a function
of $k/k_0$, for $\gamma=3$ and a set of values for the parameters $q$,
$a_0$ and $\alpha$. 
In the domain $k\ll k_0/\alpha$ where our approximations are valid, it is 
evident from the figure that, while the spectrum of the magnetic field is 
scale invariant, the spectrum of the electric field behaves as $k^{-2}$.
These are exactly the asymptotic forms~(\ref{eq:psb-bb}) and (\ref{eq:pse-bb})
that we have arrived at above.  
The question that naturally arises is whether these spectra will retain their 
form as they traverse across the bounce.

\par

\begin{figure}[!t]
\begin{center}
\includegraphics[width=12.0cm]{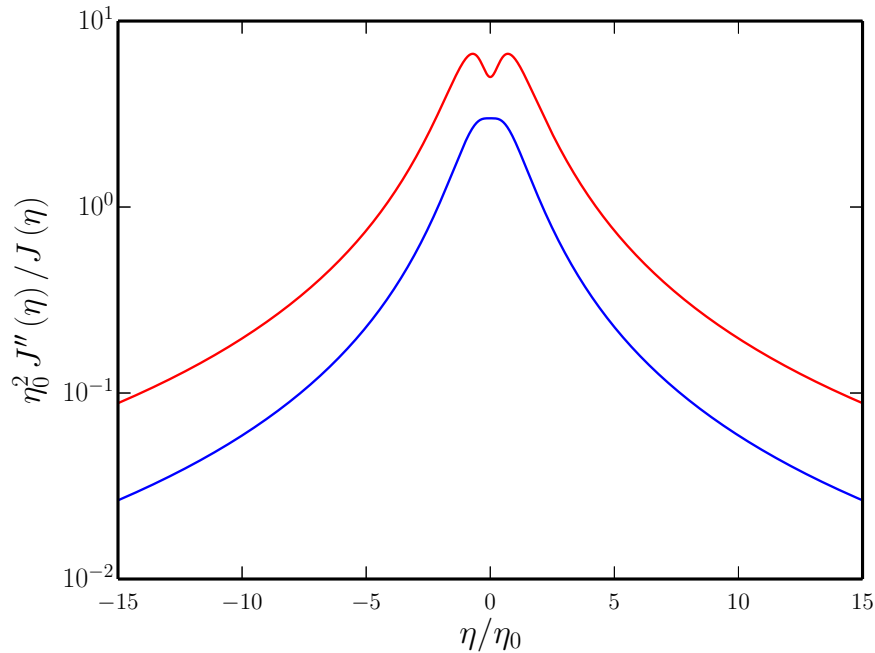}
\end{center}
\caption{The behavior of $\eta_0^2\, J''/J$, which depends only on $\eta/\eta_0$,
has been plotted for $\gamma=3$ (in blue) and $\gamma=5$ (in red). 
The figure has been plotted over a very narrow range of $\eta/\eta_0$ in order 
to illustrate the presence of a single maximum for $\gamma=3$ and two maxima 
and one minimum for $\gamma=5$.}
\label{fig:jpp-bj}
\end{figure}

Our analysis until now applies to both positive and negative values of $n$.
However, as we mentioned, we shall hereafter restrict our analysis to the cases 
wherein $n>0$.
We shall illustrate that, in such cases, one can arrive at an analytic expression 
for the electromagnetic modes even during the bounce for wavenumbers such that 
$k\ll k_0$.   
When $n>0$, $J$ grows away from the bounce and, hence, it seems natural to expect
that $J''/J$ will exhibit its maximum near the bounce.
Actually, $J''/J$ has a single maximum at the bounce for indices $n$ and $q$ 
such that $\gamma\le 3$.
One finds that, for other values of $\gamma$, there arise two maxima and a minimum 
close to the bounce.
The minimum occurs exactly at the bounce and its value proves to be $\gamma\,
k_0^2$.
These behavior are clear from Fig.~\ref{fig:jpp-bj} wherein we have plotted 
the quantity $J''/J$ for two different values of $\gamma$. 
Therefore, when $n>0$, for scales of cosmological interest such that $k\ll 
k_0$, $k^2\ll J''/J$ around the bounce.
Hence, near the bounce, we can neglect the $k^2$ term in~(\ref{eq:de-Abk-o}) 
[to be precise, we can ignore the $k^2$ term in Eq.~(\ref{eq:de-cAk})] so that 
we have 
\begin{equation}
\bA_k''+2\,\f{J'}{J}\, \bA_k'\simeq 0.
\end{equation}
This equation can be immediately integrated to yield
\begin{equation}
\bA_k'(\eta)\simeq\bA_k'(\eta_\ast)\,\f{J^2(\eta_\ast)}{J^2(\eta)},
\label{eq:bAkp-ab}
\end{equation}
where $\eta_\ast$ is a time when $k^2\ll J''/J$ before the bounce.
The above equation can be further integrated to arrive at
\begin{eqnarray}
\bA_k(\eta)
\simeq \bA_k(\eta_\ast) +\bA_k'(\eta_\ast)\,
\int\limits_{\eta_\ast}^{\eta}\d\tilde\eta\, 
\f{J^2(\eta_\ast)}{J^2(\tilde\eta)}
= \bA_k(\eta_\ast)
+\bA_k'(\eta_\ast)\,a^{2\,n}(\eta_\ast)\,
\int\limits_{\eta_\ast}^{\eta}\f{\d\tilde\eta}{a^{2\,n}(\tilde\eta)},
\label{eq:bAk-ab}
\end{eqnarray}
where we have set the constant of integration to be $\bA_k(\eta_\ast)$.
If we substitute the expression~(\ref{eq:sf}) for the scale factor, we find that 
the integral can be carried out for an arbitrary $\gamma$ to obtain that
\begin{eqnarray}
\bA_k(\eta)
\simeq \bA_k(\eta_\ast)+\bA_k'(\eta_\ast)\,\f{a^{2\,n}(\eta_\ast)}{a_0^{2\,n}}
\biggl[\eta\;{}_2F_1\l(\f{1}{2},\gamma;\f{3}{2};
-\f{\eta^2}{\eta_0^2}\r)
-\eta_\ast\;{}_2F_1\l(\f{1}{2},\gamma;\f{3}{2};
-\f{\eta_\ast^2}{\eta_0^2}\r)\biggl],
\end{eqnarray}
where ${}_2F_1(a,b;c;z)$ denotes the hypergeometric function~\cite{Mathematica8.0}.
We can now choose $\eta_\ast=-\alpha\,\eta_0$ to arrive at the behavior of 
$\bA_k(\eta)$ and $\bA_k'(\eta)$ in the second domain.
In such a case, we can make use of the solution~(\ref{eq:cA-d1}) in the first
domain to determine the values of $\bA_k(\eta_\ast)$ and $\bA_k'(\eta_\ast)$.

\par

\begin{figure}[!t]
\begin{center}
\includegraphics[width=12.0cm]{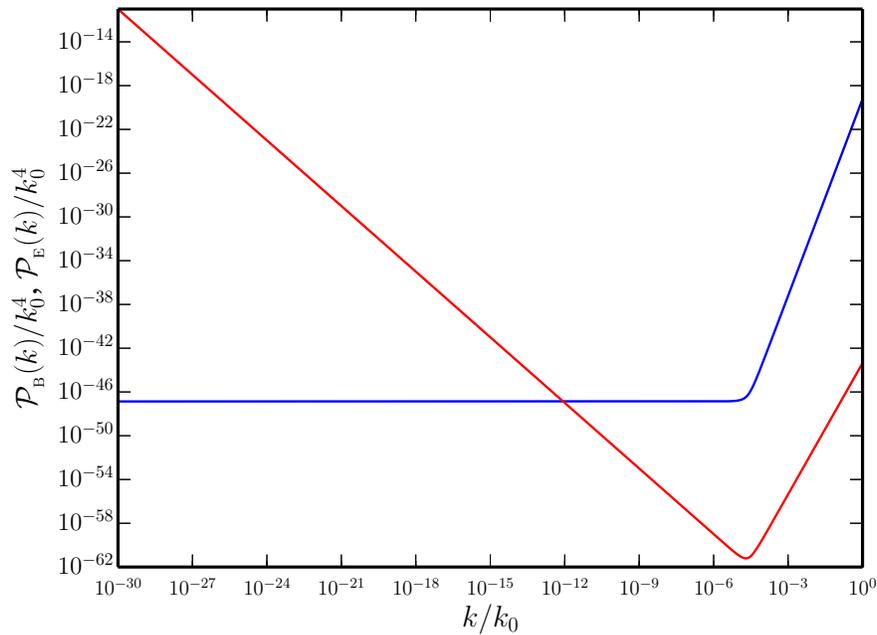}
\end{center}
\caption{The power spectra of the electric (in red) and magnetic (in blue) 
fields, evaluated at $\eta=\beta\,\eta_0$, with $\beta=10^2$, have been 
plotted for the same set of values of the parameters as in Fig.~\ref{fig:ps-bb}.
Note that the shape of the spectra generated before the bounce is retained 
for scales such that $k\ll k_0/\alpha$ even after the bounce.
We should mention that the values of the parameters we have worked with lead
to magnetic fields of observed strengths today corresponding to a few 
femto gauss.
It should also be added that the electric field dominating the strength of the
magnetic field is considered to be undesirable (see, for instance,
Ref.~\cite{Subramanian:2009fu}). 
This seems inevitable for positive $n$ that we are considering here, but it can
be, for example, circumvented by choosing $n$ to be negative (in this context, 
see Ref.~\cite{Sriramkumar:2015yza}).}
\label{fig:ps-ab}
\end{figure}

In fact, the solutions we have obtained above can be expected to be valid even 
after the bounce until the condition $k^2\ll J''/J$ is violated. 
While the bounce is symmetric, the solution $\bA_k(\eta_\ast)$ and its time
derivative $\bA_k'(\eta_\ast)$ are not symmetric~\cite{Sriramkumar:2015yza}.
Numerical analysis suggests that the analytical solutions will cease to be
valid well before the condition $k^2= J''/J$ is satisfied after the 
bounce.
For this reason, we evaluate the spectra after the bounce at $\eta=\beta\,
\eta_0$ with $\beta$ chosen to be about $10^2$. 
This choice of $\beta$ can be said to roughly correspond to the time of reheating 
after the more conventional inflationary scenario~\cite{Sriramkumar:2015yza}.
We can now evaluate the spectra {\it after}\/ the bounce at $\eta=\beta\,\eta_0$ 
using the  analytic expressions for $\bA_k$ and $\bA_k'$ we have obtained above 
[cf.~Eqs.~(\ref{eq:bAk-ab}) and~(\ref{eq:bAkp-ab})].
Recall that, while the power spectrum of the electric field depends on $\bA_k'$,
the power spectrum of the magnetic field depends on $\bA_k$.
Since $\bA_k'$ after the bounce is related to the corresponding $\bA_k'$ at the
end of the first domain only by a time dependent factor [cf.~Eq.~(\ref{eq:bAkp-ab})],
it is obvious that the shape of the electric field will not be affected by the
bounce.  
In contrast, the quantity $\bA_k$ after the bounce depends on a combination of
$\bA_k$ and $\bA_k'$ evaluated at the end of the first domain 
[cf.~Eq.~(\ref{eq:bAk-ab})].
So, it is not immediately evident that the shape of magnetic field will be 
preserved across the bounce.
In Fig.~\ref{fig:ps-ab}, we have plotted these spectra after the bounce.
Upon comparing Figs.~\ref{fig:ps-bb} and~\ref{fig:ps-ab}, it is clear that, 
while the amplitudes of the spectra change, the shapes of the spectra before 
and after the bounce are identical.


\section{The issue of backreaction}\label{sec:br}

Note that, since the electromagnetic field is a test field, the energy density 
associated with it must always remain much smaller than the energy density that 
drives the background evolution. 
However, in certain cases, it is found that the energy density associated 
with the electromagnetic field can grow and dominate the background energy 
density~\cite{Kanno:2009ei,Urban:2011bu}. 
This issue is regularly encountered in the context of inflation~\cite{Martin:2007ue}.
Such a situation is not viable and the energy density associated with the
background must be dominant at all times. 
It is therefore imperative that we examine the issue of backreaction in 
bouncing models. 
In what follows, with the analytical results at hand, we shall evaluate the 
energy density in the generated electromagnetic field and investigate the 
issue of backreaction in the bouncing scenario of our interest.

\par

\begin{figure}[!t]
\begin{center}
\includegraphics[width=12.0cm]{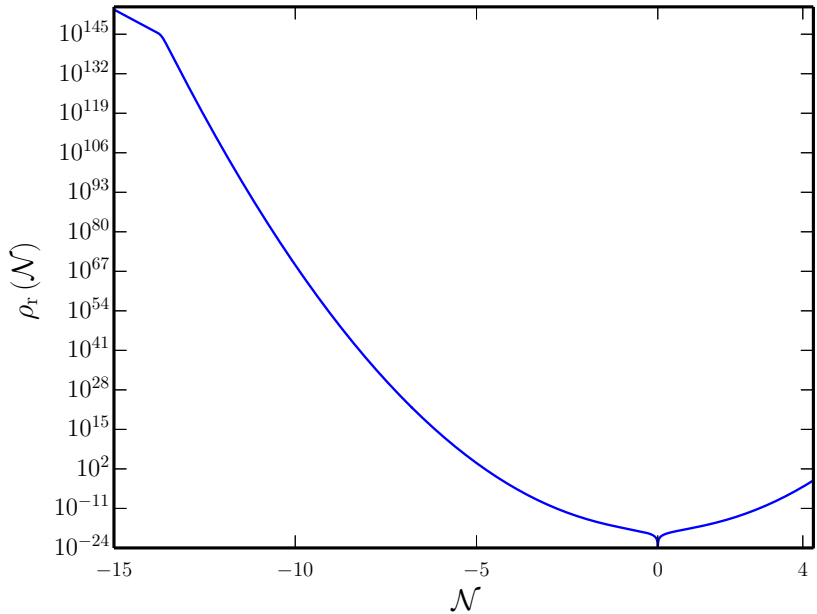}
\end{center}
\caption{The evolution of the ratio of the background energy density to the 
sum of the energy densities in the electric and magnetic fields for a given 
mode ($k=10^{-20}\,k_0$) has been plotted against e-${\cal N}$-folds.
We should mention that we have assumed the same set of values for the various
parameters as in Figs.~\ref{fig:ps-bb} and \ref{fig:ps-ab}.  
Evidently, the ratio has to remain large in order to avoid the backreaction
problem.
However, we find that the energy in the generated electromagnetic field rises 
sharply as one approaches the bounce indicating that the problem of 
backreaction is the most severe at the bounce.}
\label{fig:b-r}
\end{figure}

Using the Friedmann equation, the background energy density, say, $\rho_{\rm bg}$,
can immediately be written as
\begin{equation}
\rho_{\rm bg} = 3\,\Mpl^2\,H^2.
\end{equation}
Upon using the expression~(\ref{eq:sf}) for the scale factor, we obtain that
\begin{equation}
\rho_{\rm bg} 
= \frac{12\,\Mpl^2\,q^2\,\eta^2}{a_0^2\,\eta_0^4\,
\l(1+k_0^2\,\eta^2\r)^{2\,(q+1)}}.
\end{equation}
The energy density in a particular mode $k$ of the electromagnetic field is given
by
\begin{equation}
\rho_{_{\rm EB}}^k = \psb\l(k\r)\,+\,\pse\l(k\r).
\end{equation}
For the effects of backreaction to be negligible, the condition $\rho_{\rm bg}
>\rho_{_{\rm EB}}^k$ must be satisfied by all modes of cosmological interest 
at all times.
However, we find that this condition is violated in this scenario, particularly 
around the bounce.
To illustrate this issue,  we have plotted the ratio of 
the background energy density and the electromagnetic energy density, \viz\/ 
$\rho_{\rm r} = \rho_{\rm bg}/\rho_{_{{\rm EB}}}^k$.
We should mention that we have evaluated the quantity $\rho_{_{{\rm EB}}}^k$ from 
the analytic solutions we have obtained in the last section.
To cover a wide range in time, one often uses the e-fold as a time variable 
in the context of inflation.
In symmetric bouncing models, it proves to be convenient to use a new time 
variable ${\cal N}$ called the e-${\cal N}$-fold which is related to the scale
factor as follows~\cite{Sriramkumar:2015yza}:
\begin{equation}
a\l(\mathcal{N}\r) = a_0\,{\rm exp}\l(\mathcal{N}^2/2\r).
\end{equation}
In Fig.~\ref{fig:b-r}, we have plotted the quantity $\rho_{\rm r}$ as a function
of e-${\cal N}$-folds.
Since the Hubble parameter vanishes at the bounce, the background energy density 
also vanishes. 
Hence, any non-zero amount of electromagnetic energy density at the bounce would 
lead to a violation of the condition $\rho_{\rm r}>1$.
We find that the condition is actually violated even as one approaches the bounce 
indicating that the problem is indeed a severe one.

\par

We had mentioned earlier that no vector perturbations arise 
in the absence of vector sources.
Note that the evolution of metric vector perturbations depend on the behavior 
of the scale factor~\cite{Battefeld:2004cd}.
In contrast, the evolution of the electromagnetic modes are determined by the 
form of the coupling function~$J$.
Evidently, we do not have any vector sources in the scenario of our interest
here and, in fact, the electromagnetic modes we have considered have a quantum 
origin.
For the form of the coupling function we have assumed here [$J$ given by 
Eq.~(\ref{eq:J}), with positive~$n$], the amplitude of the generated modes 
indeed grows rapidly during the contracting phase close to the 
bounce~\cite{Sriramkumar:2015yza}.
Therefore, in this case, the issue of backreaction can be considered as a 
manifestation of the strong growth of the vector modes that are expected 
to occur as one approaches the bounce~\cite{Battefeld:2004cd}.


\section{Duality invariance}\label{sec:sidi}

The primordial scalar and tensor perturbations are governed by the so-called
Mukhanov-Sasaki equations (see, for instance, Refs.~\cite{Sriramkumar:2009kg,
Martin:2015dha}).
In these cases, it can be shown that the corresponding power spectra will 
remain invariant under a two parameter family of transformations of the 
homogeneous background quantity that determines the evolution of the 
perturbations (\viz\/ the scale factor $a$ in the case of tensor perturbations 
and a quantity often denoted as $z$ in the case of scalar 
perturbations)~\cite{Wands:1998yp}.
The new forms of the background quantities obtained as a transformation of the 
original quantities are called the dual functions.
For instance, the conventional slow roll solutions can lead to dual functions
which may be away from the slow roll limit but still produce the same power 
spectra. 
In this section, we shall extend these duality arguments to the generation 
of magnetic fields.
  
\par

The argument is in fact relatively simple.
The equation~(\ref{eq:de-cAk}) that governs the dynamics of quantity $\cA_k$
has the same form as the Mukhanov-Sasaki equations that describe the scalar
and tensor perturbations.
Note that the quantity $\cA_k$ is determined by $J''/J$.
Evidently, the solution $\cA_k$ to the differential equation can be expected
to behave in the same fashion and, hence, lead to the same power spectrum for
the magnetic field if we can construct another coupling function that leads 
to the same $J''/J$. 
Given a coupling function $J$, its dual function, say, ${\tilde J}$, which
leads to the same ${\tilde J}''/{\tilde J}$ is found to be
\begin{equation}
J(\eta) \to {\tilde J}(\eta)
= C\, J(\eta)\, \int\limits_{\eta_\ast}^{\eta}
\frac{\d{\bar \eta}}{J^2({\bar \eta})},\label{eq:jdual}
\end{equation} 
where $C$ and $\eta_\ast$ are constants. 
These constants can be suitably chosen to arrive at a physically reasonable 
form for ${\tilde J}$.

\par

Let us now construct the dual form of the coupling function~(\ref{eq:J}) that
we had considered.
The corresponding dual solution is described by the integral
\begin{equation}
{\tilde J}(\eta)
= \f{C}{J_0}\, a^n(\eta)\, 
\int\limits_{\eta_\ast}^{\eta}\frac{\d\bar{\eta}}{a^{2\,n}(\bar{\eta})}.
\end{equation}
Let us first consider the behavior at very early times when the scale 
factor~(\ref{eq:sf}) reduces to the simple power law form.
Recall that, in such a situation, the coupling function $J$ behaves as 
$J(\eta)\propto \eta^\gamma$. 
In such a case, the dual function ${\tilde J}$ can be easily evaluated
to be
\begin{equation}
{\tilde J}(\eta) 
= \f{C\, \eta^{-\gamma+1}}{-2\,\gamma+1}\,
\l(1-\f{\eta^{2\gamma-1}}{\eta_\ast^{2\gamma-1}}\r).
\end{equation}
We are specifically interested in the cases where $\gamma = 3$ and 
$\gamma = -2$, as these lead to scale invariant spectra for the 
magnetic field. 
When $\gamma = 3$, we have
\begin{equation}
{\tilde J}(\eta) 
= -\frac{C}{5\,\eta^2}\, \l(1-\f{\eta^5}{\eta_\ast^{5}}\r)
\end{equation}
and, if we set $\eta_\ast\to -\infty$, we obtain that ${\tilde J}(\eta)
\propto 1/\eta^2$.
Also, when $\gamma = -2$, we have
\begin{equation}
{\tilde J}(\eta) 
= \frac{C\,\eta^3}{5}\,\l(1-\f{\eta_\ast^5}{\eta^5}\r)
\end{equation}
and, if we can choose $\eta_\ast$ to be some large, but finite 
positive value, then at very early times, \ie\/ as $\eta\to -\infty$,
we find that ${\tilde J}(\eta)\propto \eta^3$. 
Therefore, clearly, the coupling functions corresponding to $\gamma=3$ 
and $\gamma = -2$ are dual to each other.
Given that one of these two cases leads to a scale invariant spectrum for 
the magnetic field before the bounce, their dual nature suggests that the 
other too will lead to the same spectrum, exactly as we have seen. 

\par

Let us now construct the dual form of the coupling function using the
complete scale factor~(\ref{eq:sf}), which we had used to model the 
bounce. 
On substituting the expression for the scale factor in Eq.~(\ref{eq:jdual}),
we find that we can write the dual coupling function ${\tilde J}(\eta)$ in
terms of the hypergeometric function as follows:
\begin{equation}
{\tilde J}(\eta)
= \f{C}{J_0\, a_0^n} \l(1+\frac{\eta^2}{\eta_{0}^2}\r)^{\gamma/2}\, 
\biggl[\eta\;{}_2F_1\l(\f{1}{2},\gamma;\f{3}{2};
-\f{\eta^2}{\eta_0^2}\r)
-\eta_\ast\;{}_2F_1\l(\f{1}{2},\gamma;\f{3}{2};
-\f{\eta_\ast^2}{\eta_0^2}\r)\biggl].
\end{equation}
This expression, though it is exact and is applicable to arbitrary $\gamma$, 
does not reveal the behavior of the coupling function easily. 
However, we find that for the cases corresponding to $\gamma = 3$ and $\gamma 
= -2$, ${\tilde J}(\eta)$ can be written in terms of simple functions.
When $\gamma = 3$ (say, $n=3/2$ and $q=1$), the dual form of the coupling 
function can be expressed as
\begin{eqnarray}
{\tilde J}(\eta) 
&=& \frac{C\,\eta_0}{8\,J_0\, a_0^{3/2}}\,
\l(1+\frac{\eta^2}{\eta_0^2}\r)^{3/2}\,
\Biggl[\frac{5\,(\eta/\eta_0) + 3\,(\eta/\eta_0)^3}{\l(1 + \eta^2/\eta_0^2\r)^2}
-\frac{5\,(\eta_\ast/\eta_0) + 3\,\l(\eta_\ast/\eta_0\r)^3}{\l(1+\eta_\ast^2/
\eta_0^2\r)^2} \nn\\
& & +\, 3\,\tan^{-1}\l(\frac{\eta}{\eta_0}\r)
-3\,\tan^{-1}\l(\frac{\eta_\ast}{\eta_0}\r)\Biggr].\label{eq:Jt-g3}
\end{eqnarray}
Note that the power spectrum for the electric field depends on the quantity~$J'/J$
[see Eq.~(\ref{eq:pse})]. 
Clearly, we shall require a well behaved ${\tilde J}'/{\tilde J}$ to ensure that the 
electric field evolves smoothly.
For this reason, it seems desirable to demand that the dual function ${\tilde J}$ 
does not vanish over the domain of interest.
We find that, if we set $\eta_\ast\to -\infty$, then with a suitable choice of the 
constant $C$, we can ensure that the above ${\tilde J}$ remains positive at all times. 
We also find that at early times, \ie\/ as $\eta\to -\infty$, the above ${\tilde J}$
reduces to ${\tilde J}(\eta)\propto 1/\eta^2$, as required. 
Let us now turn to the case $\gamma = -2$ (say, $n=-1$ and $q=1$).
In this case, the dual form of this coupling function is given by
\begin{eqnarray}
{\tilde J}(\eta) 
&=& \frac{C\, a_0\,\eta_0}{J_0}\,\l(1+\frac{\eta^2}{\eta_0^2}\r)^{-1}\,
\l[\f{\eta}{\eta_0}-\f{\eta_\ast}{\eta_0}+\frac{2}{3}\,
\l(\f{\eta^3}{\eta_0^3}-\f{\eta_\ast^3}{\eta_0^3}\r)
+\frac{1}{5}\, \l(\f{\eta^5}{\eta_0^5}
-\f{\eta_\ast^5}{\eta_0^5}\r)\r].
\end{eqnarray}
We find that, in such a case, if we choose $\eta_\ast$ to be a suitably 
large positive value (say, $\eta>\beta\,\eta_0$), then we can ensure that 
${\tilde J}(\eta)$ remains positive over the domain that we are 
interested in.
Also, we should point out that, at early times, \ie\/ as $\eta\to -\infty$,
the ${\tilde J}(\eta)$ above reduces ${\tilde J}(\eta)\propto \eta^3$, as 
required.

\par

\begin{figure}[!t]
\begin{center}
\includegraphics[width=12.0cm]{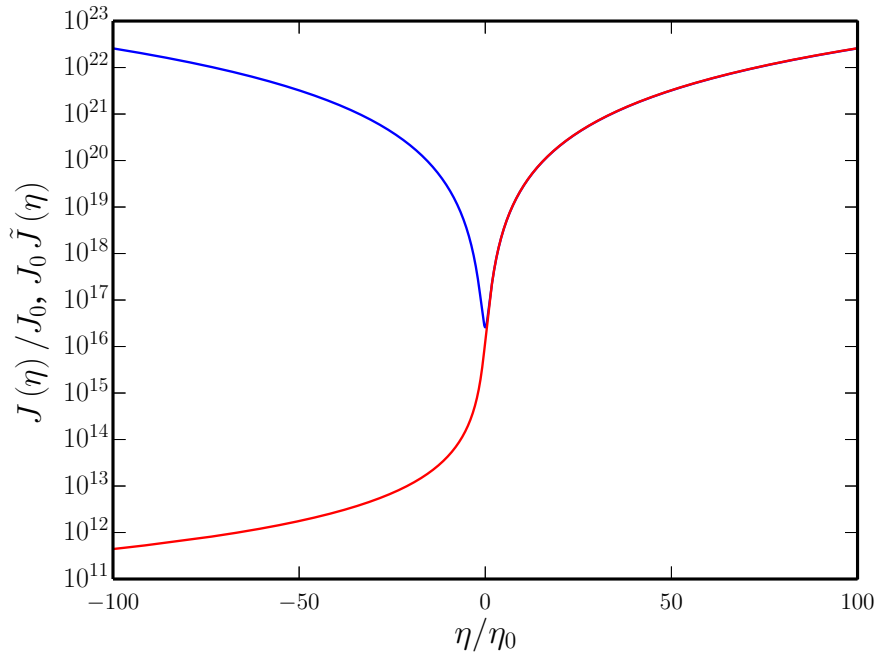}
\end{center}
\caption{The coupling function $J$ (in blue) and its dual $\tilde J$ (in red)
have been plotted as a function of $\eta/\eta_0$ for $\gamma=3$ and $\eta_\ast
\to -\infty$ [cf. Eq.~(\ref{eq:Jt-g3})]. 
Also, we have chosen the constant $C$ to be $C/k_0=5.7\times 10^{32}$ so that
the dual function ${\tilde J}$ matches the original coupling function $J$ 
after the bounce.}
\label{fig:jjdual}
\end{figure}

In Fig.~\ref{fig:jjdual}, we have plotted the coupling function $J$ and its 
dual ${\tilde J}$ for the case $\gamma=3$, with a suitable choice of the
parameters. 
Recall that our original choice for the coupling function $J$ was symmetric
about the bounce.
While the dual function ${\tilde J}$ behaves in a similar fashion as $J$ 
after the bounce (for a suitable choice of the constant $C$), we find that
the dual function behaves very differently before the bounce. 
In fact, ${\tilde J}$ is asymmetric about the bounce.
For the case of $\gamma=-2$, as we had discussed, $\eta_{\ast}$ has to be 
chosen to be a large positive value in order to ensure that ${\tilde J}$ 
does not vanish, which seems to pose difficulties for the evolution of the 
electric field. 

\section{Discussion}\label{sec:dis}

In the present work, we have {\it analytically}\/ studied the generation of 
primordial electromagnetic fields in a class of non-singular and symmetric
bouncing scenarios. 
We have assumed that the electromagnetic field is coupled non-minimally to a 
background scalar field which is expected to drive the bounce. 
Considering specific forms of the scale factor and the coupling function, we 
have arrived at analytical expressions for the power spectra for the electric 
and magnetic fields. 
We find that a scale invariant spectrum for the magnetic field arises before 
the bounce for certain values of the parameters involved, while the 
corresponding electric field spectrum has a certain power law scale dependence.
Interestingly, we have shown that, as the modes evolve across the bounce, these
shapes of the power spectra are preserved. 
However, a severe backreaction due to the generated electromagnetic fields 
seems unavoidable close to the bounce.
This issue needs to be circumvented if the scenario has to be viable. 
We have further illustrated the existence of a two parameter family of
transformations of the original coupling function under which the 
spectrum of the magnetic field remains invariant.
The dual transformation leads to asymmetric forms for the coupling function 
and it seems to be a worthwhile exercise to explore these new forms.
We are currently investigating the generation of magnetic fields in 
symmetric bounces with asymmetric coupling functions.

\par

We need to emphasize a few points at this stage of our discussion.
One may be concerned by the fact that the presence of radiation prior to the
bounce can modify the equations of motion of the electromagnetic field which 
would, in turn, affect the process of magnetogenesis.
We had described earlier as to how the class of bouncing models that we have 
considered can be driven with the aid of two fluids.
We are envisaging a situation wherein such a behavior is actually achieved with 
the help of scalar fields.
If, in addition to the scalar fields, radiation is also present before the 
bounce, its energy density can dominate close to the bounce, modifying 
the evolution of the background in the vicinity of the bounce and altering 
the form of the scale factor. 
Therefore, in our discussion, we have assumed that there is no radiation present 
before the bounce.
We believe that, after the bounce, the scalar fields driving the bounce can 
decay into radiation via some mechanism (as it occurs immediately after 
inflation) and lead to the standard radiation dominated epoch. 
However, we should add that the phenomenon of reheating in bouncing scenarios 
and its effects on the process of magnetogenesis is not yet well understood.

\par

Before concluding, we would like to make a few further 
clarifying remarks about some of the issues concerning the bouncing models 
which we had discussed in the introductory section. 
As we had described, while many bouncing models seem to lead to a large
tensor-to-scalar ratio~\cite{Allen:2004vz,Cai:2014xxa,Battefeld:2014uga,
Cai:2008qw,Cai:2014bea}, it also seems possible to construct models 
which result in scalar and tensor power spectra that are consistent with
the CMB observations~\cite{Cai:2011zx,Cai:2012va,Cai:2013kja}.
Our aim in this work was two fold.
The first was to show analytically that scale invariant magnetic fields of 
observable strengths can indeed be generated in bouncing scenarios.
The second aim was to illustrate that, just as in the case of the scalar and
the tensor perturbations, the power spectrum of the magnetic fields is
invariant under a certain duality transformation. 
Needless to say, it is important to study the generation of magnetic fields 
in those specific bouncing models which satisfy the observational constraints 
at the level of scalar and tensor power spectra.
As we had pointed out earlier, for certain values of the parameters involved, 
the form of the scale factor that we have considered here [\viz\/ Eq.~(\ref{eq:sf})]
leads to tensor power spectra that are consistent with the 
observations~\cite{Chowdhury:2015cma,Stargen:2016cft}.
Evaluating the corresponding scalar power spectra requires detailed modeling
of the source that drives the bounce.
We have been able to construct scalar field models that lead to such scale
factors.
However, we find that these scenarios require numerical efforts to arrive at 
the scalar power spectrum. 
We are currently developing codes to study these situations and compare them
with the CMB data.


\acknowledgments

DC would like to thank the Indian Institute of Technology Madras, Chennai, 
India, for financial support through half-time research assistantship.
LS wishes to thank the Indian Institute of Technology Madras, Chennai, 
India, for support through the New Faculty Seed Grant. 
RKJ would like to thank the Lundbeck foundation for financial support. 
The CP$^3$-Origins centre is partially funded by the Danish National Research
Foundation, grant number DNRF90.

\bibliographystyle{JHEP}
\bibliography{gmf-bu-as-sep-2016}

\end{document}